\documentclass[aps,pra,showpacs,superscriptaddress,preprint]{revtex4}
\usepackage{amsmath}
\usepackage{epsf}
\usepackage{epsfig}
\usepackage{bm}

\begin{document}

\title{Modified $\ell$-states of diatomic molecules subject to central potentials plus an angle-dependent potential}

\author{\small C\"uneyt Berkdemir}
\email[E-mail: ]{berkdemir@erciyes.edu.t}\affiliation{Department
of Physics, Erciyes University, 38039, Kayseri, Turkey,Turkey}
\author{Ramazan Sever}
\email[E-mail: ]{sever@metu.edu.tr}\affiliation{Department of
Physics, Middle East Technical University, 06800, Ankara,Turkey}

\date{\today}

\begin{abstract}

We present modified $\ell$-states of diatomic molecules by solving
the radial and angle-dependent parts of the Schr\"odinger equation
for central potentials, such as Morse and Kratzer, plus an exactly
solvable angle-dependent potential $V_{\theta}(\theta)/r^2$ within
the framework of the Nikiforov-Uvarov (NU) method. We emphasize
that the contribution which comes from the solution of the
Schr\"odinger equation for the angle-dependent potential modifies
the usual angular momentum quantum number $\ell$. We calculate
explicitly bound state energies of a number of neutral diatomic
molecules composed of a first-row transition metal and main-group
elements for both Morse and Kratzer potentials plus an angle-dependent potential.
\end{abstract}


\maketitle

\newpage

\section{Introduction}

In recent years, theoretical and computational studies of
molecular spectra have been one of the most valuable tools
available for studying on atoms and molecules. At its simplest
level, knowledge of spectral characteristics allows us to detect
the presence of particular characteristic and essential components
of matter. Especially, molecular spectra can be used to understand
the motion of electrons in molecules as well as the vibration and
rotation of the nuclei. The chemical interactions between atoms
and molecules assist to investigate the physical properties of
individual molecules. In the light of this knowledge, dissociation
channels \cite {1}, centrifugal distortion constants \cite {2},
semiempirical dipole moment functions \cite {3} and other data
about the rotation, vibration and electronic energy levels \cite
{4, 5, 6, 7} of diatomic molecules can be accurately determined by
using theoretical methods. Moreover, some quantum-mechanical
calculations on rotational and vibrational energy levels of
diatomic molecules have been applied to problems in molecular
physics for a number of years \cite {morales}. The modified
shifted large 1/N approach has been applied to obtain energy
levels of a rotational potential \cite {bag}, arbitrary
$\ell$-state solutions of the rotating Morse potential has been
investigated through the exact quantization rule method \cite
{wen} and other algebraic approaches and applications have been
previously applied to rotational and vibrational states of
rotating potentials \cite {lit3, tez, sever1, sever2, sever3,
sever4}.

In this study, the bound state energy levels are obtained by
solving the Schr\"odinger equation for the Morse \cite {morse} and
Kratzer \cite {kratzer} molecular potentials together with an
exactly solvable angle-dependent potential, respectively,
\begin{equation}
\label{eq2}
V_M(r,\theta)=D_e\left(e^{-2a(r-r_e)}-2e^{-a(r-r_e)}\right)+\frac{V_{\theta}(\theta)}{r^2},
\end{equation}
\begin{equation}
\label{eq3}
V_K(r,\theta)=-D_e+D_e\left(\frac{r-r_e}{r}\right)^2+\frac{V_{\theta}(\theta)}{r^2},
\end{equation}
where subscripts $M$ and $K$ indicate the Morse and Kratzer
potentials, respectively. $a$ controls the width of the potential
and $r_e$ is the equilibrium internuclear distance. The quantity
$D_e$ is the electronic (or spectroscopic) dissociation energy of
the diatomic molecule and it differs slightly from the chemical
dissociation energy $D_0$, i.e., $D_0=D_e-\hbar \omega_e/2$, where
$\omega_e$ is called harmonic vibrational parameter \cite {brans,
ogilvie}. Moreover, the minimum value of $V_{M,K}(r)$ at $r=r_e$
belongs to $D_e$. The second term in the right-hand side of
Eq.(\ref{eq2}) or Eq.(\ref{eq3}) represents an angle-dependent
potential and its uncovered form is given as follows

\begin{equation}
\label{eq4}
V_{\theta}(\theta)=\frac{\hbar^2}{2\mu}\left(\frac{A}{sin^2
\theta}+\frac{B}{cos^2 \theta}\right).
\end{equation}

$A$ and $B$ in Eq.(\ref{eq4}) are fixed constants or parameters
obtained by some fitting procedure which is based on experimental
or theoretical results; it is important to emphasize that they
cannot depend on the angle $\theta$. The factor $\hbar^2/2\mu$ is
introduced in view of future convenience. The potential given in
Eq.(\ref{eq4}) has been introduced for the first time by Makarov
$et.al$ \cite {maka} classifying some non-central potential
systems. The Schr\"odinger equation for this type of
angle-dependent potential can be exactly solved to obtain the
bound state energies of a diatomic molecule. It is well-known that
the problem of exact solution of the Schr\"odinger equation for a
number of special potentials has been a line of great interest in
some quantum mechanical applications. The solution of this
equation for some potential has been made by applying some
analytical methods. One of these methods is developed by Nikiforov
and Uvarov \cite {nikiforov} as a new approach to the theory of
special functions. They succeeded in obtaining an unified integral
representation for functions of hypergeometric type. This type of
hypergeometric equation with an appropriate coordinate
transformation is given as follows

\begin{equation}
\label{eq5} \psi ^{\prime \prime }(s)+\frac{\stackrel{\sim }{\tau
}(s)}{\sigma (s)}\psi ^{\prime }(s)+\frac{\stackrel{\sim }{\sigma
}(s)}{\sigma ^{2}(s)}\psi (s)=0
\end{equation}

where $~\sigma (s)$ and $~\stackrel{\sim }{\sigma }(s)~$ are
polynomials, at most second$-$degree, and $~\stackrel{\sim }{\tau
}(s)~$ is a first$-$degree polynomial. The general view point of
this paper is to present an analytical solution of the
angle-dependent part of the Schr\"odinger equation for an exactly
solvable angle-dependent potential $V_{\theta}(\theta)/r^2$ and
also to obtain modified $\ell$ states of diatomic molecules. The
solution method developed by Nikiforov and Uvarov is used for
solving the Schr\"odinger equation. The angle-dependent part of
the Schr\"odinger equation is investigated in detail to derive
some analytical result and the solution of the radial part of the
associated equation for the Morse and Kratzer potentials is
extracted from the papers published previously \cite {berk1,
berk2}. The modified $\ell$ state expressions for the Morse and
Kratzer potentials are obtained by connecting the results of the
angle-dependent part with the radial one. The modified $\ell$
states of a number of neutral diatomic molecules composed of a
first-row transition metal and main-group elements are calculated
for both Morse and Kratzer potentials with an angle-dependent
potential.

\section{Separating variables of the Schr\"odinger equation in spherical coordinates}

The starting point of this section is to separate the
Schr\"odinger equation in spherical coordinates for a diatomic
molecule represented by a rotating potential model. After
separating the center of mass motion, the eigenvalue equation for
a rotating motion in spherical coordinates is solved by using the
NU method and the energy levels of the discrete spectrum are
obtained for several diatomic molecules. In spherical coordinates,
the Schr\"odinger equation is written as follows:

\begin{eqnarray}
\label{eq6}
\left\{-\frac{\hbar^2}{2\mu}\left[\frac{1}{r^2}\frac{\partial}{\partial
r}\left(r^2\frac{\partial}{\partial r}\right) +\frac{1}{r^2
sin\theta}\frac{\partial}{\partial \theta}\left(sin
\theta\frac{\partial}{\partial \theta}\right)+\frac{1}{r^2
sin^2\theta}{\frac{\partial^2}{\partial\varphi^2}}\right]\right\}\Psi_{n\ell
m}(\textbf{r})\nonumber\\
+V(\textbf{r})\Psi_{n\ell m}(\textbf{r})=E\Psi_{n\ell
m}(\textbf{r}).
\end{eqnarray}

The energy $E$ in Eq.(\ref{eq6}) is real and it is either discrete
for bound states ($E<0$) or continuous for scattering states
($E>0$). Introducing a new variable $x=cos^2 \theta$,
Eq.(\ref{eq6}) can be explicitly turned into the more useful one:

\begin{eqnarray}
\label{eq7} \left\{\frac{1}{r^2}\frac{\partial}{\partial
r}\left(r^2\frac{\partial}{\partial
r}\right)+\frac{1}{r^2}\left[4x(1-x)\frac{\partial^2}{\partial
x^2}+2(1-3x)\frac{\partial}{\partial x} +
\frac{1}{1-x}\frac{\partial^2}{\partial
\varphi^2}\right]\right\}\Psi_{n\ell m}(\textbf{r})\nonumber\\
+\frac{2\mu}{\hbar^2}(E-V(\textbf{r}))\Psi_{n\ell
m}(\textbf{r})=0.
\end{eqnarray}

Consequently, this equation is separable for a potential of the
following form,

\begin{equation}
\label{eq8}
V(\textbf{r})=V_{M,K}(r)+\frac{1}{r^2}\left[V_{\theta}(x)+\frac{1}{1-x}V_{\varphi}(\varphi)\right].
\end{equation}

If we write the wave function as $\Psi_{n\ell
m}(\textbf{r})=r^{-1}R_{n\ell}(r)\Theta_{\ell
m}(\theta)\Phi_m(\varphi)$, then the wave equation in
Eq.(\ref{eq7}) with the potential in Eq.(\ref{eq8}) is separated
to a set of second-order differential equations in all three
coordinates as follows:

\begin{equation}
\label{eq9}
\left(\frac{d^2}{dr^2}-\frac{E_{\theta}}{r^2}+\frac{2\mu}{\hbar^2}(E-V_{M,K}(r))\right)R_{n\ell}(r)=0,
\end{equation}
\begin{equation}
\label{eq10}
\left(4x(1-x)\frac{d^2}{dx^2}+2(1-3x)\frac{d}{dx}-\frac{E_{\varphi}}{1-x}+E_{\theta}-\frac{2\mu}{\hbar^2}V_{\theta}(x)\right)\Theta_{\ell
m}(x)=0,
\end{equation}

\begin{equation}
\label{eq11}
\left(\frac{d^2}{d\varphi^2}-\frac{2\mu}{\hbar^2}V_{\varphi}(\varphi)+E_{\varphi}\right)\Phi_m(\varphi)=0,
\end{equation}

where $E_{\varphi}$ and $E_{\theta}$ are the separation constants,
which are real and dimensionless. Since the wave function
$\Psi_{n\ell m}(\textbf{r})$ must be finite in all space for the
bound states, the boundary conditions for Eq.(\ref{eq9}) require
$R_{n\ell}(0)=0$ and the square-integrability of $R_{n\ell}(r)$ on
$(0,\infty)$, which implies that $R_{n\ell}(\infty)=0$. The finite
solutions for $\Theta_{\ell m}(\theta)$ in the range $0\leq \theta
\leq \pi$ are able to map into a differential equation of
hypergeometric type. Moreover, the boundary conditions for
Eq.(\ref{eq11}) must be $\Phi_m(\varphi+2\pi)=\Phi_m(\varphi)$. If
the azimuthal-dependent potential part $V_{\varphi}(\varphi)$ is
set up to zero, then the normalized solution of Eq.(\ref{eq11})
that satisfies the boundary conditions becomes

\begin{equation}
\label{eq12}
\Phi_m(\varphi)=\frac{1}{\sqrt{2\pi}}e^{im\varphi},~~~~~~~~m=0,\pm
1,\pm 2,...,
\end{equation}
where one of the separation constants $E_{\varphi}$ represents
$m^2$, i.e., $E_{\varphi}=m^2$.

\subsection{The Solution of Eq.(\ref{eq10})}

It is well-known that the solution of the radial part of the
Schr\"odinger equation gives eigenvalues and eigenfunctions for a
particle moving within the interaction potentials. However, the
solution of the angle-dependent part of the corresponding equation
does not depends on eigenvalues presented in the solution of the
radial part explicitly. It only exhibits a parameter relationship
between contribution constants which come from the
$\theta$-dependent part of the potential. Such a relationship can
be expressed by solving Eq.(\ref{eq10}) in terms of $E_{\theta}$.
Eq.(\ref{eq10}) can then be rewritten in the following form by
introducing an exactly solvable angle-dependent potential given in
Eq.(\ref{eq4}),

\begin{equation}
\label{eq13}
\left(4x(1-x)\frac{d^2}{dx^2}+2(1-3x)\frac{d}{dx}-\frac{E_{\varphi}}{1-x}+
E_{\theta}-\left(\frac{A}{1-x}+\frac{B}{x}\right)\right)\Theta_{\ell
m}(x)=0.
\end{equation}

An arrangement of the above equation turns to a convenient form to
make a comparison with the main equation of the NU method given in
Eq.(\ref{eq5});

\begin{eqnarray}
\label{eq14} \frac{d^2\Theta_{\ell
m}(x)}{dx^2}+\frac{(1-3x)}{2x(1-x)}\frac{d\Theta_{\ell
m}(x)}{dx}+\frac{1}{\left[2x(1-x)\right]^2}\times\nonumber \\
\left(-E_{\theta}x^2+x(E_{\theta}-\widetilde{A}+B)-B\right)\Theta_{\ell
m}(x)=0,
\end{eqnarray}

where $\widetilde{A}=m^2+A$ (keeping in mind the selection of
$E_{\varphi}=m^2$). Having compared Eq.(\ref{eq14}) with
Eq.(\ref{eq5}), the following polynomial equalities are obtained
immediately

\begin{equation}
\label{eq15} \widetilde{\tau}=1-3x,
\end{equation}
\begin{equation}
\label{eq16} \sigma=2x(1-x),
\end{equation}

\begin{equation}
\label{eq17}
\widetilde{\sigma}=-E_{\theta}x^2+x(E_{\theta}-\widetilde{A}+B)-B.
\end{equation}

In the next step, the basic solution procedure of the NU method
given in Ref.\cite {berk1} will be followed to find a solution of
Eq.(\ref{eq14}) in terms of $E_{\theta}$. If polynomials given in
Eqs.(\ref{eq15})-(\ref{eq17}) are substituted into Eq.(6) of
Ref.\cite {berk1}, $\pi$ function is obtained as follows

\begin{eqnarray}
\label{eq18}
\pi=\frac{1-x}{2}\pm\frac{1}{2}\sqrt{x^2(4E_{\theta}-8k+1)-x(4E_{\theta}-4\widetilde{A}+4B-8k+2)+1+4B}.
\end{eqnarray}

The simplest form of $\pi$ can be written
\begin{equation}
\label{eq19} \pi=\frac{1-x}{2}\pm \frac{1}{2}\sqrt{\alpha
x^2-\beta x+\gamma},
\end{equation}
where $\alpha=4E_{\theta}-8k+1$,
$\beta=4E_{\theta}-4\widetilde{A}+4B-8k+2$ and $\gamma=1+4B$. The
possible solutions according to the plus and minus signs of
Eq.(\ref{eq19}) depend on the parameter $k$ within the square root
sign. The expression under the square root has to be the square of
a polynomial, since $\pi$ is a polynomial of degree at most 1. To
satisfy this condition, the discriminant of the expression within
the square root must be set up to zero, i.e., $\Delta
=\beta^2-4\alpha \gamma=0$. This identity leads to

\begin{equation}
\label{eq20}
(4E_{\theta}-4\widetilde{A}+4B-8k+2)^2-4(4E_{\theta}-8k+1)(1+4B)=0,
\end{equation}
and a second-order equation related to $k$ is originated as
follows

\begin{equation}
\label{eq21}
4k^2+4k(\widetilde{A}+B-E_{\theta})+(\widetilde{A}-B)^2-2E_{\theta}(\widetilde{A}+B)
+E_{\theta}^2-\widetilde{A}=0.
\end{equation}
Hence, the double roots of $k$ are derived as
\begin{equation}
\label{eq22} k_{1,2}=-\frac{(\widetilde{A}+B-E_{\theta})}{2}\pm
\frac{1}{2}\sqrt{\widetilde{A}(1+4B)}.
\end{equation}
Substituting $k_{1,2}$ into Eq.(\ref{eq18}), the four possible
solutions of $\pi$ are obtained
\begin{equation}
\label{eq23} \pi = \frac{1-x}{2}\pm
\frac{1}{2}\left\{\begin{array}{cc}
\left[\left(2\sqrt{\widetilde{A}}-\sqrt{1+4B}\right)x+\sqrt{1+4B}\right],\\
\hskip 0.5cm \mbox{for} \hskip 0.5cm
k_1=-\frac{(\widetilde{A}+B-E_{\theta})}{2}
+\frac{1}{2}\sqrt{\widetilde{A}(1+4B)}\\\\
\left[\left(2\sqrt{\widetilde{A}}+\sqrt{1+4B}\right)x-\sqrt{1+4B}\right],\\
\hskip 0.5cm \mbox{for} \hskip 0.5cm
k_2=-\frac{(\widetilde{A}+B-E_{\theta})}{2}-\frac{1}{2}\sqrt{\widetilde{A}(1+4B)}\\
\end{array}\right.
\end{equation}

where $k_{1,2}$ is determined by means of the same procedure as in
Ref.\cite {berk1}. We have to choose one of the four possible
forms of $\pi$ to obtain the bound state solutions. Therefore, its
most suitable form is established by
$\pi=\frac{1-x}{2}-\frac{1}{2}\left[\left(2\sqrt{\widetilde{A}}+\sqrt{1+4B}\right)x-\sqrt{1+4B}\right]$
for
$k_2=-\frac{(\widetilde{A}+B-E_{\theta})}{2}-\frac{1}{2}\sqrt{\widetilde{A}(1+4B)}$.
The main requirement in the selection of this form is to find the
negative derivative of $\tau(s)$ given by Eq.(9) of Ref.\cite
{berk1}. In that case, $\tau(s)$ and $\tau^{\prime}(s)$ are
obtained, respectively,

\begin{eqnarray}
\label{eq24}
\tau(s)=1+\sqrt{1+4B}-x\left(4+2\sqrt{\widetilde{A}}+\sqrt{1+4B}\right), \nonumber \\
\tau^{\prime}(s)=-\left(4+2\sqrt{\widetilde{A}}+\sqrt{1+4B}\right)
<0~.
\end{eqnarray}

Another major polynomials given in the basic solution procedure of
the NU method are $\lambda$ and $\lambda _{\widetilde{n}}$ \cite
{nikiforov}. Both polynomials can be connected with each other by
means of Eq.(7) and Eq.(8) of Ref.\cite {berk1}. Hence, a
polynomial of degree $\widetilde{n}$ is found by using $\lambda
_{\widetilde{n}}=-\widetilde{n}\tau ^{\prime
}-\frac{\widetilde{n}(\widetilde{n}-1)}{2}\sigma ^{\prime \prime
}$;
\begin{equation}
\label{eq25}
\lambda_{\widetilde{n}}=2\widetilde{n}^2+2\widetilde{n}+2\widetilde{n}\sqrt{\widetilde{A}}
+\widetilde{n}\sqrt{1+4B},~~~(\widetilde{n}=0,1,2,...)
\end{equation}
taking $\sigma^{\prime \prime}=-4$. Moreover, $\lambda$ is
obtained from $k_2+\pi^{\prime}$;
\begin{equation}
\label{eq26}
\lambda=-\frac{1}{2}\sqrt{1+4B}\left(1+\sqrt{\widetilde{A}}\right)
-\frac{1}{2}\left(\widetilde{A}+B-E_{\theta}+1\right)-\sqrt{\widetilde{A}}
\end{equation}

After comparing Eq.(\ref{eq25}) with Eq.(\ref{eq26}) and also
making some arrangements on the comparison, the separation
constant $E_{\theta}$ is obtained as follows

\begin{equation}
\label{eq27}
\left(2\widetilde{n}+\sqrt{\widetilde{A}}\right)^2+2\sqrt{\widetilde{A}}+\sqrt{1+4B}
+\left(2\widetilde{n}+\sqrt{\widetilde{A}}\right)\sqrt{1+4B}+(1+B)=E_{\theta}.
\end{equation}

It is very useful to prepare
$\widetilde{\ell}(\widetilde{\ell}+1)$ as a new presentation
instead of $E_{\theta}$. In this case, Eq.(\ref{eq27}) turns to

\begin{equation}
\label{eq28}
\left(1/2+2\widetilde{n}+\sqrt{\widetilde{A}}+\sqrt{1/4+B}\right)
\left(1/2+2\widetilde{n}+\sqrt{\widetilde{A}}+\sqrt{1/4+B}+1\right)=\widetilde{\ell}(\widetilde{\ell}
+1),
\end{equation}
and it becomes in terms of $\widetilde{\ell}$
\begin{equation}
\label{eq29}
\widetilde{\ell}=\left(1/2+2\widetilde{n}+\sqrt{\widetilde{A}}+\sqrt{1/4+B}\right).
\end{equation}

The term $\widetilde{\ell}$ in Eq.(\ref{eq29}) can be named the
"modified" orbital angular momentum, since the contribution which
comes from the angle-dependent potential damages the usual orbital
angular momentum $\ell$. Moreover, the result obtained in
Eq.(\ref{eq29}) is in agreement with results on the more involved
case of Ref.\cite {kibler}. In the limiting case $B=0$, the factor
$\sqrt{1/4+B}$ in Eq.(\ref{eq29}) should be replaced by $\pm 1/2$
so that Eq.(\ref{eq29}) turns into $\nu + \sqrt{\widetilde{A}}$,
where $\nu=1+2\widetilde{n}$ for the odd functional solution or
$\nu=2\widetilde{n}$ for the even functional solution \cite
{quesne}. The parameter $\widetilde{\ell}$ does not need to be
integer. However, the difference between the parameter
$\widetilde{\ell}$ and the square root terms in Eq.(\ref{eq29})
have to be integer;

\begin{equation}
\label{eq30}
\widetilde{n}=\frac{1}{2}\left\{\widetilde{\ell}-\left(1/2+\sqrt{\widetilde{A}}+\sqrt{1/4+B}\right)\right\},
~~~~\widetilde{n}=0,1,2,...
\end{equation}
where $\widetilde{n}$ corresponds to the number of quanta for
oscillations.

\subsection{The Solution of Eq.(\ref{eq9})}

It is remarkable that the radial equation in Eq.(\ref{eq9}) is
independent of the angle-dependent term given in Eqs.(\ref{eq2})
and (\ref{eq3}) for the Morse and Kratzer cases, respectively.
Eq.(\ref{eq9}) is exactly soluble by means of the NU method.
However, some caution must be observed especially on the solution
of the Morse potential since the exponential nature of the Morse
potential and the radial behavior of the centrifugal kinetic
energy term do not allow for solving the Schr\"odinger equation
simultaneously. In the case of Kratzer potential, no caution is
necessary when considering the Kratzer potential together with the
centrifugal term since both terms shows the radial behaviors. In
the following subsections, the solution of both potentials is
briefly investigated by using the NU method.

\subsubsection{The Morse case}

Adopting the Morse potential to Eq.(\ref{eq9}), the radial
Schr\"odinger equation turns into the following form
\begin{equation}
\label{eq31}
\left(\frac{d^2}{dr^2}-\frac{E_{\theta}}{r^2}+\frac{2\mu}{\hbar^2}
\left(E-D_e\left(e^{-2a(r-r_e)}-2e^{-a(r-r_e)}\right)\right)\right)R_{n\ell}(r)=0.
\end{equation}
Disadvantage of Eq.(\ref{eq31}) is that analytical solutions
cannot be found because of the centrifugal kinetic energy term of
the potential proportional to $E_{\theta}/r^2$ is included into
the radial Schr\"odinger equation. In order to obtain an
analytical solution of Eq.(\ref{eq31}), the term $E_{\theta}/r^2$
has to be approximated to the exponential one. Using an accurate
approximate treatment suggested by Pekeris \cite {pekeris}, this
term can be translated into the following form
\begin{equation}
\label{eq32} \frac{E_{\theta}}{r^2}\cong
\frac{E_{\theta}}{r_e^2}\left(D_0+D_1e^{-ar_e x}+D_2e^{-2ar_e
x}\right),
\end{equation}
where $x$ is a coordinate transformation represented by
$(r-r_e)/r_e$ and $D_i$ is the coefficients which are given in
Eq.(18) of Ref.\cite{berk1} (i=0,1,2). Substituting
Eq.(\ref{eq32}) into Eq.(\ref{eq31}) and using a new variable of
the form $s=e^{-ar_ex}$, the resulting Schr\"odinger equation
becomes
\begin{equation}
\label{eq33}
\frac{d^2R_{n\ell}(s)}{ds^2}+\frac{1}{s}\frac{dR_{n\ell}(s)}{ds}
+\frac{1}{s^2}\left[-\varepsilon_1^2
+\varepsilon_2s-\varepsilon_3s^2\right]R_{n\ell}(s)=0,
\end{equation}
where
$-\varepsilon_1^2=2\mu\left(E-\frac{E_{\theta}D_0}{r_e^2}\right)/\hbar^2a^2$,
$\varepsilon_2=2\mu\left(2D_e-\frac{E_{\theta}D_1}{r_e^2}\right)/\hbar^2a^2$
and
$\varepsilon_3=2\mu\left(D_e-\frac{E_{\theta}D_2}{r_e^2}\right)/\hbar^2a^2$.
Comparing this equation with that of Eq.(21) of Ref.\cite{berk1}
and following the solution steps of the NU method, the energy
spectrum according to the quantum numbers $n$, $\widetilde{n}$ and
$m$ is obtained as
\begin{eqnarray}
\label{eq34} E_{n\widetilde{n}m} =\frac{\hbar^2E_{\theta}}{2\mu
r_e^2}\left(1-\frac{3}{ar_e}+\frac{3}{a^2r_e^2}\right)-\frac{\hbar^2a^2}{2\mu}
\left[C_{\widetilde{n}m}-\left(n+\frac{1}{2}\right)\right]^2,
\end{eqnarray}
where
\begin{eqnarray} \label{eq35} C_{\widetilde{n}m}=\frac{1}{\sqrt{\frac{2\mu a^2D_e}{\hbar^2}
+\frac{a^2E_{\theta}D_2}{r_e^2}}}\left[\frac{2\mu
D_e}{\hbar^2}-\frac{E_{\theta}}{r_e^2}\left(\frac{2}{ar_e}-\frac{3}{a^2r_e^2}\right)\right],
\end{eqnarray}
and $E_{\theta}$ is given by Eq.(\ref{eq28}), keeping in mind
$\widetilde{A}=m^2 +A$. The highest vibrational quantum number
$n_{max}$ can be directly estimated from the condition
$dE_{n\widetilde{n}m}/dn=0$;
\begin{eqnarray} \label{eq36}
n_{max}=C_{\widetilde{n}m}-\frac{1}{2}.
\end{eqnarray}
$n_{max}$ is generally limited to obtain the number of bound
states in the case of the Morse potential and its maximum value
depends on the potential parameters of a given diatomic molecule
as well as the quantum numbers $\widetilde{n}$ and $m$.

\subsubsection{The Kratzer case}

Among many two-particle interaction models, one of the most
interesting potential types is the Kratzer potential because it
can be exactly solved for the general case of rotation states
different from zero. The first term on the right-hand side of
Eq.(\ref{eq4}) is the central Kratzer potential and the radial
part of the Schr\"{o}dinger equation in the presence of this
potential can be written as follows, recalling Eq.(\ref{eq9}),
\begin{equation}
\label{eq37}
\left(\frac{d^2}{dr^2}-\frac{E_{\theta}}{r^2}+\frac{2\mu}{\hbar^2}
\left[E+D_e-D_e\left(\frac{r-r_e}{r}\right)^2\right]\right)R_{n\ell}(r)=0.
\end{equation}
Using the transformation $s\rightarrow r/r_e$ and letting the
dimensionless notations
\begin{equation}
\label{eq38} -\varepsilon_1^2=\frac{2\mu r_e^2E}{\hbar^2},~~~~
\varepsilon_2=\frac{4\mu
D_er_e^2}{\hbar^2}~~~~\varepsilon_3=E_{\theta}+\frac{2\mu
D_er_e^2}{\hbar^2},
\end{equation}
Eq.(\ref{eq37}) can be rewritten in a simple form as follows
\begin{equation}
\label{eq39}
\frac{d^2R_{n\ell}(s)}{ds^2}+\frac{1}{s^2}\left(-\varepsilon_1^2s^2+\varepsilon_2
s-\varepsilon_3\right)R_{n\ell}(s)=0.
\end{equation}
The complete solution of Eq.(\ref{eq39}) by means of the NU method
can be found in Ref.\cite {berk2}, after having made of some
notation setting. Hence, the energy spectrum with respect to the
quantum numbers $n$, $\widetilde{n}$ and $m$ is obtained as
\begin{equation}
\label{eq40} E_{n\widetilde{n}m}
=-\frac{\hbar^2}{2\mu}\left[\left(\frac{4\mu
D_er_e}{\hbar^2}\right)^2
\left(1+2n+\sqrt{1+4D_{\widetilde{n}m}}~\right)^{-2}\right],
\end{equation}
where
\begin{eqnarray}
\label{eq41} D_{\widetilde{n}m}=\frac{2\mu
D_er_e^2}{\hbar^2}+\left(1/2+2\widetilde{n}+\sqrt{m^2+A}+\sqrt{1/4+B}\right)\nonumber\\
\times\left(1/2+2\widetilde{n}+\sqrt{m^2+A}+\sqrt{1/4+B}+1\right).
\end{eqnarray}

The derivative of Eq.(\ref{eq41}) according to $n$ gives the
maximum vibrational quantum number $n_{max}$ in the case of
Kratzer potential;
\begin{eqnarray} \label{eq42}
\frac{dE_{n\widetilde{n}m}}{dn}=\frac{\frac{8\mu
D_e^2r_e^2}{\hbar^2}}{\left(1+2n_{max}+\sqrt{1+4D_{\widetilde{n}m}}~\right)^{3}}=0.
\end{eqnarray}
The condition which requires to satisfy the equality on the
right-hand side of Eq.(\ref{eq42}) is that $n_{max}$ must be
supported by an infinite number of vibrational levels.

\subsection{Remarks and calculations for the modified $\ell$ states}

In order to discuss the behavior of energy spectrums of a diatomic
molecule when the values of quantum numbers $n$, $\widetilde{n}$
and $m$ differ, it is very useful to select some diatomic
molecules composed of a first-row transition metal and main-group
elements (H-F). One or two of these molecules are the first-row
transition metal hydrides such as ScH, TiH, VH, CrH and MnH
\cite{112}. Transition metal hydrides are chemical compounds
formed when hydrogen gas reacts with transition metal atoms. These
are of considerable importance in chemical synthesis as
intermediates and in solid matrix samples for infrared
spectroscopic study. Another diatomic molecule containing the
transition metal element copper (Cu) and the main group element
lithium (Li) is CuLi, which elucidates the nature of the bonding
in mixed transition metal lithides \cite {179}. Presently the
transition metal carbide molecules such as TiC and NiC represent a
very active field of research, especially due to the desire for a
quantitative understanding of their chemical bonds \cite
{185,191}. Moreover, diatomic scandium nitride molecule ScN has
excellent physical properties of high temperature stability as
well as electronic transport properties, which are typical of
transition metal nitride \cite {195}. Furthermore, the scandium
fluoride molecule ScF is the best studied transition metal halide
and it has been fairly well characterized \cite {287}. Diatomic
molecules which consist of transition metal and main group
elements are challenging theoretically and computationally, but
recent advancements in computational methods have made such
molecules more accessible to investigations. Their spectroscopic
parameters have been accurately determined by using $ab$-$initio$
calculations. One of these calculations is called the
multi-configuration self-consistent field (MCSCF) and it seems
qualitatively correct. In Table 1, the spectroscopic parameters of
the above mentioned diatomic molecules are summarized using MCSCF
results \cite {har}. However, choice of the parameter $a$ is not a
simple issue. Solution of the Schr\"odinger equation for the Morse
potential gives the following well-known relation (see p.132 of
Ref.\cite {ogilvie});
\begin{equation}\label{eq44}
a=\frac{\omega_e}{2r_e\sqrt{B_eD_e}},
\end{equation}
where $B_e=\frac{hc}{8\pi^2\mu c^2r_e^2}$. Notice that the
parameter $a$ is used to calculate the energy spectrum of the
Morse potential. Another considerable effort for the Morse
potential is that the highest vibrational quantum number $n_{max}$
changes according to the spectroscopic parameters of diatomic
molecules as well as the parameters $\widetilde{n}$, $m$, $A$ and
$B$, keeping in mind Eq.(\ref{eq36}). As an example, the value of
$n_{max}$ for ScH is 20 in the fixed values of $A=1$ and $B=9$ and
under the conditions of $\widetilde{n}\leq 10$ and $m\leq 10$. The
values of $n_{max}$ for TiH, VH, CrH, MnH, CuLi, TiC, NiC, ScN and
ScF molecules given in Table 1 are aligned 20, 20, 17, 14, 70, 71,
50, 100 and 131, respectively, in the same values of parameters
and conditions.

To calculate the bound state energies of diatomic molecules given
in Table 1, Eqs.(\ref{eq34}) and (\ref{eq40}) must be recalled for
the Morse and Kratzer cases, respectively. Taken into account
spectroscopic parameters of diatomic molecules and arbitrary
values of $A$ and $B$, the bound state energies can be compared
for both potentials. This type of comparison is given in Table 2.
As can be seen from Table 2, when parameters $A$ and $B$ are fixed
to $1$ and $1$, respectively, for different values of $n$,
$\widetilde{n}$ and $m$, the bound state energies become lower
than that of other values. A comparison of $A=1$ and $B=9$ with
$A=9$ and $B=1$ shows that the bound state energies obtained for
$A=1$ and $B=9$ are a little smaller than the energies obtained
for $A=9$ and $B=1$ in small values of $n$, $\widetilde{n}$ and
$m$, especially 0 and 1. For large values of the quantum numbers,
the bound state energies obtained for $A=9$ and $B=1$ tend to
become more separately spaced than the energies obtained for $A=1$
and $B=9$.

\section{Conclusions}

An interesting extension of this work is to study the effect of an
angle-dependent potential to the Morse and Kratzer potentials and
to examine the partial changes on the usual $\ell$ states. The
analysis presented in this work suggests that the bound state
energies of diatomic molecules depend on the quantum numbers $n$,
$\widetilde{n}$, $m$ and also the parameters $A$ and $B$.
Moreover, the energy spectrum obtained in Eq.(\ref{eq34}) is an
approximate description of the quantum aspects of diatomic
molecules for the Morse potential together with angle-dependent
potential while the spectrum obtained in Eq.(\ref{eq40}) is a
complete description for the Kratzer potential together with
angle-dependent potential. Furthermore, the solution procedure
presented in this paper is also systematical and efficient for
solving the angle-dependent part of the Schr\"odinger equation.

\section{Acknowledgments}

The authors acknowledge partially support from the Scientific and
Technological Research Council of Turkey (TUBITAK). Moreover, CB
acknowledges the support of Science Foundation of Erciyes
University.

\newpage

\begin{table}
\caption{Spectroscopic parameters and reduced masses for some
diatomic molecules composed of a first-row transition metal and
main-group elements (H-F). The complete list of this table can be
found from Ref.\cite {har}.}
\begin{tabular}{ccccccc}\hline\hline
\hspace*{0.2cm} Molecule & \hspace*{0.2cm} $D_e$~(eV)&
\hspace*{0.2cm} $r_e$~($\AA$)& \hspace*{0.2cm}
$\omega_e$~($cm^{-1}$) & \hspace*{0.2cm} $a$~($\AA^{-1}$) & \hspace*{0.2cm} $\mu$~(a.m.u) & Reference\\
\hline
ScH & 2.25 & 1.776 & 1572 & 1.41113 & 0.986040 &\cite {112}\\
TiH & 2.05 & 1.781 & 1407 & 1.32408 & 0.987371 &\cite {112}\\
VH & 2.33 & 1.719 & 1635 & 1.44370 & 0.988005 &\cite {112}\\
CrH & 2.13 & 1.694 & 1647 & 1.52179 & 0.988976 &\cite {112}\\
MnH & 1.67 & 1.753 & 1530 & 1.59737 & 0.989984 &\cite {112}\\
CuLi & 1.74 & 2.310 & 392 & 1.00818 & 6.259494 &\cite {179}\\
TiC & 2.66 & 1.790 & 592 & 1.52550 & 9.606079 &\cite {185}\\
NiC & 2.76 & 1.621 & 874 & 2.25297 & 9.974265 &\cite {191}\\
ScN & 4.56 & 1.768 & 726 & 1.50680 & 10.682771 &\cite {195}\\
ScF & 5.85 & 1.794 & 713  & 1.46102 & 13.358942 &\cite {287}\\
\hline
\end{tabular}
\end{table}

\newpage

\begin{table}
\caption{The variation of bound state energies (in eV) for various
values of $n$, $\widetilde{n}$, $m$, $A$ and $B$}
\begin{tabular}{cccc|ccc|ccc}
\hline\hline &  &  &  &
& Morse Potential &  &  & Kratzer Potential &  \\
\hline
Molecule & n & $\widetilde{n}$ & m & $A=1$ & $A=1$ & $A=9$ & $A=1$ & $A=1$ & $A=9$  \\
&  &  &  & $B=9$ & $B=1$
&  $B=1$ & $B=9$ & $B=1$ & $B=1$ \\
\hline
ScH & 0 & 0 & 0 & -2.13697 & -2.14733 & -2.13645 &-2.19509 & -2.20526 &-2.19459  \\
~~~~ & 1 & 1 & 0 & -1.93560 & -1.95052 & -1.93490 &-2.10731  & -2.12154 & -2.10665 \\
~~~~ & 3 & 2 & 1 & -1.56637 & -1.58578 & -1.56832 & -1.95151 & -1.96918 & -1.95327 \\
~~~~ & 3 & 3 & 2 & -1.53010 & -1.55582 & -1.53794 & -1.91920 &-1.94202  &-1.92611  \\
~~~~ & 5 & 4 & 3 & -1.17869 & -1.20852 & -1.19224 & -1.76519 & -1.79029 & -1.77652\\
~~~~ & 5 & 5 & 4 & -1.12534 & -1.16101 & -1.14521 &-1.72169  &
-1.75058 & -1.73769 \\ \hline
TiH & 0 & 0 & 0 & -1.94719 & -1.95748 & -1.94668 &-1.99713 & -2.00721 & -1.99663 \\
~~~~ & 1 & 1 & 0 &-1.76538 & -1.78019 & -1.76469 &-1.91331  &-1.92738  & -1.91266  \\
~~~~ & 3 & 2 & 1 & -1.43216 & -1.45141 & -1.43409 & -1.76537 &-1.78272 & -1.76710 \\
~~~~ & 3 & 3 & 2 & -1.39624 & -1.42170 & -1.40400 & -1.73372 & -1.75607 & -1.74048  \\
~~~~ & 5 & 4 & 3 & -1.07703 & -1.10651 & -1.09041 & -1.58802 & -1.61244 & -1.59903 \\
~~~~ & 5 & 5 & 4 & -1.02440 & -1.05957 & -1.04398 & -1.54584 &
-1.57383 & -1.56133 \\ \hline
VH & 0 & 0 & 0 & -2.21203 & -2.22307 & -2.21148 &-2.27210 & -2.28293 & -2.27156 \\
~~~~ & 1 & 1 & 0 & -2.00226 & -2.01816 & -2.00153 & -2.17978 & -2.19492 & -2.17908 \\
~~~~ & 3 & 2 & 1 & -1.61793 & -1.63859 & -1.62000 & -2.01624 & -2.03499 & -2.01810 \\
~~~~ & 3 & 3 & 2 & -1.57935 & -1.60671 & -1.58769 & -1.98197 & -1.07718 & -1.98929 \\
~~~~ & 5 & 4 & 3 & -1.21343 & -1.24513 & -1.22783 & -1.82048 & -1.84705 & -1.83247\\
~~~~ & 5 & 5 & 4 & -1.15677 & -1.19465 & -1.17787 &-1.77450  &
-1.80503 & -1.79140 \\ \hline
CrH & 0 & 0 & 0 & -2.01092 & -2.02226 & -2.01036 & -2.07289& -2.08400 & -2.07234 \\
~~~~ & 1 & 1 & 0 & -1.80031 & -1.81659 & -1.79956 & -1.96202 & -1.98219 & -1.96113 \\
~~~~ & 3 & 2 & 1 & -1.41769 & -1.43870 & -1.41980 & -1.8247 & -1.84369& -1.82659 \\
~~~~ & 3 & 3 & 2 & -1.37846 & -1.40627 & -1.38694 & -1.79010 & -1.81452 &-1.79749 \\
~~~~ & 5 & 4 & 3 & -1.01871 & -1.05067 & -1.03322 & -1.63474 & -1.66128 & -1.64671\\
~~~~ & 5 & 5 & 4 & -0.96162 & -0.99978 & -0.98287 & -1.58901 &
-1.61934 & -1.60579 \\ \hline
MnH & 0 & 0 & 0 & -1.55956 & -1.57012 & -1.55904 & -1.61987& -1.63018 & -1.61936 \\
~~~~ & 1 & 1 & 0 & -1.36574 & -1.38081 & -1.36504 & -1.54231 &-1.55656  & -1.54165 \\
~~~~ & 3 & 2 & 1 & -1.01943 & -1.03864 & -1.02136 & -1.40762 & -1.42489& -1.40934 \\
~~~~ & 3 & 3 & 2 & -0.98358 & -1.00900 & -0.99133 & -1.37632 & -1.39839 & -1.38298 \\
~~~~ & 5 & 4 & 3 & -0.66694 & -0.69570 & -0.68000 & -1.24530 & -1.26892 & -1.25593\\
~~~~ & 5 & 5 & 4 & -0.61561 & -0.64992 & -0.63471 & -1.20492 &
-1.23166 & -1.21969 \\ \hline \hline
\end{tabular}
\end{table}

\newpage

\begin{tabular}{cccc|ccc|ccc}
\hline\hline &  &  &  &
& Morse Potential &  &  & Kratzer Potential &  \\
\hline
Molecule & n & $\widetilde{n}$ & m & $A=1$ & $A=1$ & $A=9$ & $A=1$ & $A=1$ & $A=9$  \\
&  &  &  & $B=9$ & $B=1$
&  $B=1$ & $B=9$ & $B=1$ & $B=1$ \\
\hline
CuLi & 0 & 0 & 0 & -1.71422 & -1.71519 & -1.71417 & -1.72804& -1.72901 & -1.72799 \\
~~~~ & 1 & 1 & 0 & -1.66482 & -1.66626 & -1.66475 & -1.70610 &-1.70752  & -1.70603 \\
~~~~ & 3 & 2 & 1 & -1.56867 & -1.57064 & -1.56886 & -1.66395 & -1.66586 & -1.66414 \\
~~~~ & 3 & 3 & 2 & -1.56497 & -1.56759 & -1.56577 & -1.66038 & -1.66291 & -1.66115 \\
~~~~ & 5 & 4 & 3 & -1.46933 & -1.47256 & -1.47080 & -1.61763 & -1.62070 & -1.61903\\
~~~~ & 5 & 5 & 4 & -1.46352 & -1.46741 & -1.46569 & -1.61213 &
-1.61581 & -1.61419  \\ \hline
TiC & 0 & 0 & 0 & -2.62172 & -2.62278 & -2.62167 &-2.61837 & -2.61941 & -2.61832 \\
~~~~ & 1 & 1 & 0 & -2.54773 & -2.54930 & -2.54766 & -2.59060 & -2.59213 & -2.59052 \\
~~~~ & 3 & 2 & 1 & -2.40344 & -2.40559 & -2.40366 & -2.53690 &-2.53898 & -2.53711  \\
~~~~ & 3 & 3 & 2 & -2.39942 & -2.40228 & -2.40029 & -2.53301 & -2.53577 & -2.53385 \\
~~~~ & 5 & 4 & 3 & -2.25676 & -2.26028 & -2.25836 &-2.47861  & -2.48196  & -2.48013\\
~~~~ & 5 & 5 & 4 & -2.25042 & -2.25467 & -2.25279 & -2.47257 &
-2.47661 & -2.47482 \\ \hline
NiC & 0 & 0 & 0 & -2.70409 & -2.70533 & -2.70402 & -2.74321& -2.74445 & -2.74315 \\
~~~~ & 1 & 1 & 0 & -2.59598 & -2.59781 & -2.59589 & -2.71217 &-2.71398  & -2.71208 \\
~~~ & 3 & 2 & 1 & -2.38692 & -2.38942 & -2.38717 & -2.65228 & -2.65473& -2.65252 \\
~~~~ & 3 & 3 & 2 & -2.38223 & -2.38556 & -2.38325 & -2.64768 & -2.65094 & -2.64868  \\
~~~~ & 5 & 4 & 3 & -2.17893 & -2.18301 & -2.18079 & -2.58698 & -2.59093 & -2.58878\\
~~~~ & 5 & 5 & 4 & -2.17157 & -2.17650 & -2.17432 & -2.57987 &
-2.58463 & -2.58252 \\ \hline
ScN & 0 & 0 & 0 & -4.51353 & -4.51451 & -4.51348 & -4.54157& -4.54255 & -4.54152 \\
~~~~ & 1 & 1 & 0 & -4.42292 & -4.42437 & -4.42285 & -4.50668 & -4.50812 & -4.50662 \\
~~~ & 3 & 2 & 1 & -4.24493 & -4.24693 & -4.24513 & -4.43862 &-4.44058 & -4.43881 \\
~~~~ & 3 & 3 & 2 & -4.24119 & -4.24385 & -4.24200 & -4.43494 & -4.43755 & -4.43573 \\
~~~~ & 5 & 4 & 3 & -4.06444 & -4.06774 & -4.06594 &-4.36611  & -4.36932 & -4.36757\\
~~~~ & 5 & 5 & 4 & -4.05850 & -4.06248 & -4.06072 & -4.36033
&-4.36420 & -4.36249 \\ \hline
ScF & 0 & 0 & 0 &-5.80466 & -5.80542 & -5.80462 &-5.83194  &-5.83270 & -5.83190  \\
~~~~ & 1 & 1 & 0 & -5.71576 & -5.71689 & -5.71571 & -5.79735 & -5.79848 & -5.79730 \\
~~~~ & 3 & 2 & 1 & -5.54042 & -5.54198 & -5.54057 &-5.72950  & -5.73104 & -5.72965 \\
~~~~ & 3 & 3 & 2 & -5.53749 & -5.53957 & -5.53813 & -5.72661 & -5.72866 & -5.72723  \\
~~~~ & 5 & 4 & 3 &-5.36298 & -5.36556 & -5.36415 & -5.65811 & -5.66064 & -5.65926 \\
~~~~ & 5 & 5 & 4 & -5.35832 & -5.36144 & -5.36006 & -5.65355 &
-5.65660 & -5.65525 \\ \hline
\end{tabular}

\newpage
\begin{figure}[htbp]
\includegraphics[scale=1.0]{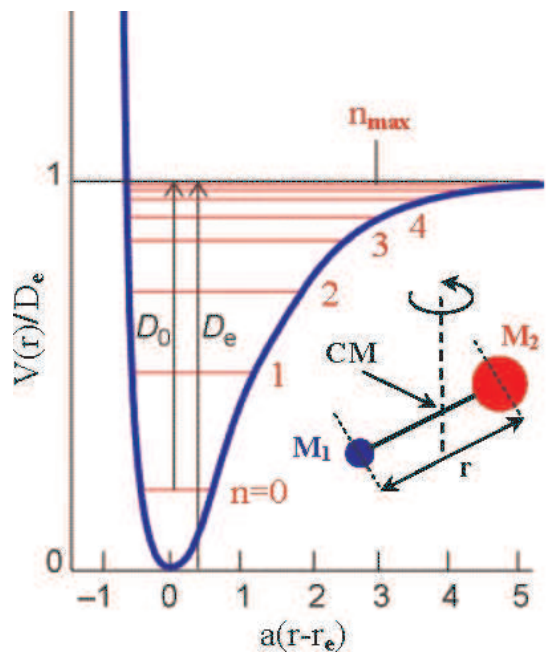}
\caption{Representative vibrational energy levels and rotation of a
diatomic molecule. $n$ is the vibration quantum number and $D_0$ is
the chemical dissociation energy of the lowest ($n = 0$) vibrational
level.  The internuclear distance $r$ is representatively shown in
the right-hand sight of figure.}\label{eps1}
\end{figure}

\end{document}